\documentclass[aps,prx,twocolumn,english,superscriptaddress,floatfix,longbibliography]{revtex4-2}
\usepackage{graphicx}
\usepackage{dcolumn}
\usepackage{bm}
\usepackage{amsmath}
\usepackage[english]{babel}
\usepackage[utf8]{inputenc}
\usepackage{color}
\usepackage{amssymb}

\usepackage{palatino}
\usepackage[hidelinks]{hyperref}

\usepackage{braket}

\begin{document}

\title{Satellite-assisted Entanglement Distribution with High-Dimensional Photonic Encoding}

\author{V. Domínguez Tubío}
\affiliation{QuTech, Delft University of Technology, 2628 CJ Delft, The Netherlands}
\affiliation{Kavli Institute of Nanoscience, Delft University of Technology, 2628 CJ, Delft, The Netherlands}
\author{M.C. Dijksman}
\affiliation{QuTech, Delft University of Technology, 2628 CJ Delft, The Netherlands}
\affiliation{Kavli Institute of Nanoscience, Delft University of Technology, 2628 CJ, Delft, The Netherlands}
\author{J. Borregaard}
\affiliation{Department of Physics, Harvard University, Cambridge, Massachusetts 02138, USA}
\affiliation{Lightsynq Technologies Inc., Brighton, Massachusetts 02135, USA}

\date{\today}

\begin{abstract}

Satellite-assisted entanglement distribution is a promising approach for realizing long-range quantum networking. However, the limited coherence time of existing quantum memories makes it challenging to obtain multiple event-ready entangled pairs between ground stations since one pair decoheres before the successful distribution of another. We demonstrate how this can be circumvented by pairing existing satellite-compatible spontaneous parametric down conversion (SPDC) sources with qudit-compatible quantum memories on ground. By operating the SPDC source as a source of time-bin encoded photonic qudits, simultaneous distribution of multiple entangled pairs between the ground stations can be achieved at a significantly higher rate than if the SPDC sources was operated as a source of photonic qubits. We find that for achievable coherence times of several seconds and demonstrated satellite performances from the Micius satellite, the qudit operation leads to several orders of magnitude faster distribution rates than the qubit-based operation when more than one event-ready high-quality (Bell pair fidelity $\geq0.95$) entangled pair is desired. To ensure high-quality entanglement distribution, we consider multiplexed quantum memory operation storage and, in the qubit case, we also consider storage cutoff times.  

\end{abstract}

\maketitle

\section{Introduction}

Encoding information into quantum systems leads to fundamentally new capabilities for data processing and transmission~\cite{Wehner2018QuantumAhead}. Quantum key distribution (QKD)~\cite{Pironio2009Device-independentAttacks, Scarani2009TheDistribution} and blind quantum computing~\cite{Fitzsimons2017PrivateProtocols,VanMeter2016TheComputing} can enable information-theoretic security in communication and cloud quantum computing provided that entanglement can be established between the communicating parties. The standard approach for entanglement distribution involves sending photons through optical fibres. However, for long-distance links, satellite-assisted free space links is a promising near-term approach that circumvents the need of complex quantum repeaters to compensate transmission loss~\cite{forges2023,Liorni_2021,gundogan2021,Wallnofer_2022,Khatri_2021,Boone_2015}. For example, QKD has already been demonstrated with the Micius satellite as a trusted node for distances ranging from 500 to 700 km \cite{qkd_trusted_node,vienna-beiging-qkd}. In a more advanced approach, entanglement-based QKD was achieved without relying on a trusted node, spanning distances of up to 1,200 km ~\cite{yin_entanglement-based_2020}, which is well outside the reach of any current fibre-based approach.

To unlock the full potential of entanglement-based quantum networking, the distribution of multiple high-quality entangled pairs will be necessary. This allows for combating noise and imperfections through entanglement purification~\cite{Krastanov2019,pattison2024} as well as more advanced applications such as distributed or blind quantum computing~\cite{Ramette2024npj,wei2024}. In conventional photonic qubit-based approaches, multiple photonic qubit pairs are distributed sequentially~\cite{Kalb2017EntanglementNodes} or in parallel~\cite{Jiang2009QuantumEncoding,Munro2015InsideRepeaters} and rely on a register of quantum memories to store the successful pairs until all required entangled pairs are distributed. This puts daunting requirements on the coherence time of the quantum memories, which will be proportional to the inverse of the transmission probability of the photonic link~\cite{Zheng2022EntanglementQudits}. To address this challenge, it has been proposed to use high-dimensional photonic qudit encoding to simultaneously entangle multiple pairs~\cite{Zheng2022EntanglementQudits,Xie2021QuantumMultiplexing}. However, the compatibility of such protocols with realistic satellite-based entanglement sources, as well as their comparison with other qubit-compatible techniques like temporal multiplexing~\cite{collins2007,Simon2010,Largo2023} and storage-time cutoffs~\cite{Li2021,inesta2023}, remains largely unexplored.  

In this work, we propose a promising scheme for long-distance entanglement distribution where a satellite-based SPDC source is operated as an approximate source of photonic qudits to entangle two ground-based quantum memory registers. The qudit operation of the SPDC sources does not require any significant upgrade of the quantum hardware on the satellite, which we assume have similar characteristics as the Micius satellite~\cite{yin_entanglement-based_2020}. Qudit compatible quantum memories are required for the ground nodes, which can be realized with existing hardware based on cavity-coupled diamond color centers or atomic qubits~\cite{Zheng2022EntanglementQudits,Bhaskar2020ExperimentalCommunication,Knaut2024,Reiserer2022}.  

We compare the performance of the SPDC qudit operation with the standard qubit operation considering multiplexed memory operation in both cases and storage cutoff times for the qubit one. For the former, we assume the quantum memories can store more qubit pairs than required, which, in some cases, can result in an effectively higher distribution rate. For the latter, we assume that an entangled pair is discarded if it has been stored longer than a certain cut-off time since the quality of the entanglement will have degraded too much from decoherence.  By operating with a qudit dimension that corresponds to the desired number of event-ready entangled pairs, the memories only need to store the entangled pairs for the duration of classical communication between two ground stations. This leads to several orders of magnitude faster distribution rates for high-fidelity entanglement distribution (Bell pair fidelity $\geq 0.90$) over distances ranging from hundreds to over a thousand kilometers compared to the qubit operation for similar parameters of the source, quantum memories, and satellite-link. Our results demonstrate the practical advantage of high-dimensional photonic encoding for long-range entanglement distribution with existing satellite-based quantum hardware and near-term ground-based quantum memories.     

\section{Model}

\begin{figure}[htp]
   \centering
    \includegraphics[width=1.0\linewidth]{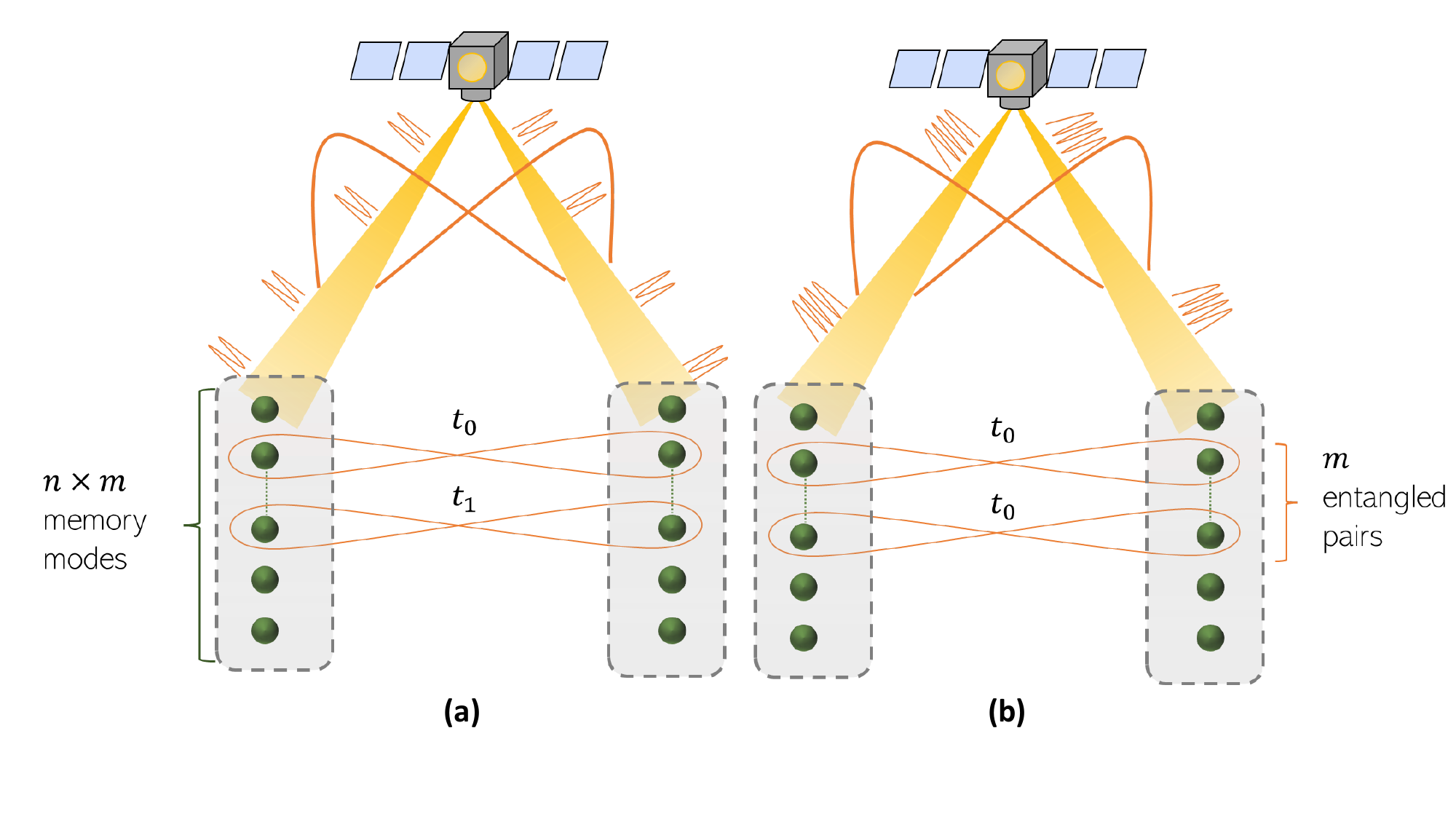}
    \caption{\textbf{Experimental setup.} We consider a down-link scenario where the entangled photons are transmitted from the satellite to the ground stations. We also consider multiplexing, where we have more memory modes, $n\times m$ than pairs we want to entangle. Here $m$ is the desired number of pairs and $n$ is the multiplexing factor. \textbf{(a) Qubit mode of operation.} The source fires entangled photons encoded in a two-dimensional time-bin encoding. If both photons arrive successfully at their ground stations, they are stored in the quantum memories, depicted as the first successful pair at $t_0$. The other successful pair arrives at a later time, $t_1$. \textbf{(b) Qudit mode of operation.} The source fires entangled photons in a $2^m$-dimensional time-bin encoding. When one pair of entangled photons arrives successfully at both ground stations, we simultaneously obtain the desired number of Bell pairs.}
    
    \label{fig:fig_1}
\end{figure}

We consider a downlink setup where photons are sent from the satellite to the ground stations, as sketched in Fig.~\ref{fig:fig_1}. The goal is to establish $m$ event-ready high-fidelity entangled pairs between two ground stations meaning that at a certain point in time, we have $m$ entangled pairs simultaneously available. We consider two distinct modes of operation.

In the first qubit mode of operation, shown in Fig.~\ref{fig:fig_1}(a), approximate entangled photonic qubit pairs are generated from an SPDC source and transmitted to the ground stations. The photonic qubits are assumed to be encoded in the time-bin basis consisting of an early or late time bin. Upon successful arrival at their respective nodes, the photons are stored in time-bin compatible quantum memories in a heralded manner. This can e.g. be achieved through spin-dependent reflection of cavity-coupled diamond defect centers or atomic quantum memories as we will detail below. We will consider cases where the number of available memory modes is equal or higher than the required number of entangled pairs. Furthermore, we include the possibility of operating with a cutoff time such that pairs that have been stored for too long and correspondingly been subject to too much decoherence will be discarded. 

In the second qudit mode of operation, depicted in Fig.\ref{fig:fig_1}(b), the SPDC source generates approximate entangled photonic qudit pairs encoded in a high-dimensional time-bin basis. We note that this merely corresponds to viewing the consecutive emission of $2^{m-1}$ approximate qubit pairs as an approximate $2^m$ dimensional qudit as we detail below. Thus, the operation of the SPDC source is in many ways exactly the same as in the qubit mode of operation. The ground-based memories will, however, need to store high-dimensional qudit states of dimension $2^m$ in a heralded manner. Similarly to the qubit case, multiple cavity-coupled diamond defect centers or atomic qubits can be used to achieve this. Once successfully heralded, a photonic qudit pair generates $m$ Bell pairs between the ground stations simultaneously. We will always consider the case where the qudit dimension matches the desired number of Bell pairs. For this reason, we do not consider cutoff strategies for the qudit mode of operation but still include the possibility of having more memory modes than the required number of Bell pairs to increase the distribution rate through multiplexing.

\subsection{Entangled photon pair source}

We consider an SPDC source as our probabilistic source of entangled photonic pairs. A pump laser is used to drive a nonlinear optical crystal to emit time-correlated signal and idler photon pairs with a certain probability~\cite{yin_entanglement-based_2020,Yin2017Satellite-basedKilometers,Fedrizzi2009High-fidelityChannel}. To generate time-bin encoded states, we consider a pulsed operation corresponding to well-defined time bins of length $\tau_{\text{rep}}$. Two consecutive pulses generates an approximate time-bin qubit pair and, in general, $2^m$ consecutive pulses generates an approximate $2^m$-dimensional entangled qudit pair of the form:

\begin{eqnarray}
    |\Psi\rangle_{\text{A,B}}&=&\bigotimes_{l=0}^{2^\text{m}-1}\sqrt{1-\lambda^2}\sum_{n=0}^\infty\lambda^n|n_l,n_l\rangle_{A,B} \nonumber\\
    &&\approx(1-\lambda^2)^{2^{m-1}}\bigl[|0,0\rangle_{A,B} +\lambda\sum_{l=0}^{2^m1}|1_l,1_l\rangle_{A,B}+ \nonumber \\
    && + \lambda^2 \sum_{l=0}^{2^m-1}\sum_{j\geq l}^{2^m-1}|1_l1_j,1_l1_j\rangle_{A,B} + O(\lambda^3) \bigr],
    \label{eq:Full_qudit_state}
\end{eqnarray}

where the subscripts $A$ and $B$ correspond to two different spatial modes directed toward the ground stations of Alice and Bob respectively, and the subscripts $l,j$ enumerate the time bins that contain photons. In other words, the state $\ket{1_l1_j,1_l1_j}_{A,B}$ denotes a state with photon pairs (in spatial mode $A$ and $B$) in the $l$'th and $j$'th timebin. The desired term in Eq.~(\ref{eq:Full_qudit_state}) is $\sum_{l=0}^{2^m1}|1_l,1_l\rangle_{A,B}$, which corresponds to a (not normalized) maximally entangled qudit pair of dimension $2^m$. 

We assume the source operates in a weak-pump regime, where the squeezing parameter is much less than one ($\lambda \ll 1$). Lowering the squeezing parameter decreases the probability of emitting a photon pair, yet improves the fidelity by reducing multi-photon events ($\propto \lambda^2$). In other words, there is a trade-off between the rate and the fidelity of entanglement which depends on the dimension of the entangled pair.

\subsection{Photon transmission}

Photon loss during transmission to the two ground stations will change the received state from the one emitted by the SPDC source. We model this by passing the state in Eq.~(\ref{eq:Full_qudit_state}) through fictitious beam splitters in each spatial mode where one output mode of the beam splitters corresponds to the loss while the other is the transmitted mode. As we detail below, we assume that the storage of the received photons in the quantum memories is heralded by the detection of the photon. Any loss and inefficiency in the memory storage process can be directly included in the effective transmission probability, $p_T$, of the fictitious beam splitter together with the the free space transmission probability. The free space transmission probability is derived using the model in Ref.~\cite{tubio2024satelliteassistedquantumcommunicationsingle}, which includes divergence, atmospheric absorption and pointing jitter. Additionally, we model false detections assuming that there is a probability $p_{\text{\text{dark}}}$ that a detector clicks despite no photon was received. We assume that this results in a stored state with zero fidelity with the desired entangled Bell pairs to have a lower bound on the Bell pair fidelity.

\subsection{Quantum memory operation}

Upon arrival at the ground stations, the incoming photonic state is mapped to a (multi-)qubit state in the qubit registers of the ground stations. We assume the availability of qubit-photon controlled gates, which makes the transformation
\begin{equation}
\ket{\text{vac}}_{\text{ph}}\ket{0}\to\ket{\text{vac}}_{\text{ph}}\ket{0}, \qquad \ket{1}_{\text{ph}}\ket{0}\to\ket{1}_{\text{ph}}\ket{1},
\end{equation}
where $\ket{\text{vac}}_{\text{ph}}$ ($\ket{1}_{\text{ph}}$) denotes the absence (presence) of a photon in a particular mode. Such gates can be implemented with existing hardware of cavity-coupled diamond color centers~\cite{Bhaskar2020ExperimentalCommunication,Knaut2024} or atomic qubits~\cite{Ritter2012AnCavities,Reiserer2022} and can be used to store $2^m$ dimensional photonic qudits in $m$ qubits in a heralded manner with the protocol of  Ref.~\cite{Zheng2022EntanglementQudits}. Here, we sketch the main steps of this protocol and note that it also works for photonic qubits ($m=1$). 

To simplify the explanation, we consider only the desired component of the photon state in Eq.~(\ref{eq:Full_qudit_state}):
\begin{equation}
    |\Psi\rangle = \frac{1}{\sqrt{2^m}}\sum_{l=0}^{2^m-1}|l\rangle_{A,ph}|l\rangle_{B,ph}.
    \label{eq:desired_state_2}
\end{equation}

\begin{figure}[htp]
   \centering
    \includegraphics[width=1.0\linewidth]{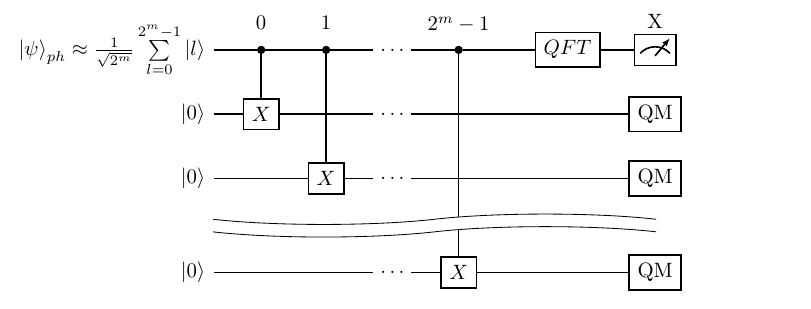}
    \caption{\textbf{Qudit state mapping to multi-qubit states.} Quantum circuit illustrating the qudit memory protocol at the ground stations. The control nodes represent optical switches that route the incoming time-bin pulses to the corresponding quantum memories based on the binary encoding of the time-bin. At the memories, a qubit-photon controlled note gate is performed which entangles the qubit state of the memories with the photon.  After the interaction, the incoming photon is measured in the Fourier basis, which heralds the successful simultaneous storage of multiple Bell pairs.}
    
    \label{fig:fig_2}

\end{figure}

At both Alice and Bobs side, the corresponding qudit interacts with $m$ cavity-coupled qubit memories. Which of the $m$ qubit memories that the photon interacts with depends on the specific time-bin and is controlled by photonic switching. The switching is programmed according to the binary representation of the time-bins (see Fig.~\ref{fig:fig_2}). Consequently, the decimal time-bin encoding of the $2^m$-dimensional qudits in Eq.~(\ref{eq:desired_state_2}) will be transformed into the corresponding binary encoding in $m$ qubits. A $2^m$-dimensional qudit state represented by $|l\rangle$ $(l=0,1,\dots,2^m-1)$ can be changed to binary encoding by the following conversion: $[l]_{10}=[l_{m-1}\cdots l_k\cdots l_1l_0]_2$ ($l_i \in \{0,1\}$)~\cite{Zheng2022EntanglementQudits}. For example, the number $5_{10}$ is mapped to 101$_2$ in binary encoding. 

The $m$-qubit register is initialized in the $|0\rangle^{\otimes m}$ state, and the incoming photon flips the qubits corresponding to its binary-encoded state. If $l_i=1$ the $i$th qubit of the register will flip from $|0\rangle$ to $|1\rangle$. If $l_i=0$ the qubit remains in state $|0\rangle$. After interaction at both ground stations, the resulting state of the qudit and qubit memory registers is:

\begin{equation} \label{eq:qudit-memory1}
    |\psi\rangle = \frac{1}{\sqrt{2^m}}\sum_{l=0}^{2^m-1}|l\rangle_{A,ph}|l\rangle_{B,ph}|l_2\rangle_A|l_2\rangle_B,
\end{equation}

where $|l_2\rangle$ represents the state of the multi-qubit register in binary encoding. The successful storage of the qudit information in the qubit registers is completed by detecting the photonic modes in a generalized X-basis such that the information about which time-bin the photons were in is erased. Such a measurement can be implemented by applying local quantum Fourier transforms (QFT) to the photonic modes at the ground stations, which can be implemented with optical delay lines, beam splitters and phase shifters as outlined in Refs.~\cite{Zheng2022EntanglementQudits,Xiaoyu2024}. It corresponds to projecting the photonic modes onto basis states of the form 
\begin{equation}
|X_k\rangle=\frac{1}{\sqrt{2^m}}\sum_{l=0}^{2^m-1}e^{2i\pi l k/2^m}|l\rangle,
\end{equation}
where $k\in\{0,1,\dots,2^m-1\}$. The measurement also heralds the successful transfer of the qudit information to the qubit register i.e. that a photon was received. Up to local phases determined by the measurement outcomes at Bob and Alice, the qubit register state following detection will be:

\begin{equation}
    |\psi\rangle = \frac{1}{\sqrt{2}}(|0,0\rangle_{A,B}+ |1,1\rangle_{A,B})^{\otimes m}.
\end{equation}

To confirm the success of both sides, the measurement outcomes must be communicated between Alice and Bob. During this time, referred to as the heralding time, $\tau_h$, the stored qubits experience decoherence. Additionally, in the qubit case, when more than one event-ready entangled pair is desired, the already entangled pairs need to wait until all desired pairs have been successfully entangled during which they also suffer from decoherence. We model this as a depolarizing channel where the quantum state decays to a mixed state over time ($t$), with a decoherence probability of $1-\text{exp}(-t/\tau_{coh})$. Here, $\tau_{coh}$ is the coherence time of the memories.

Besides the limited decoherence time of the qubit registers, additional errors such as photon loss and imperfect spin-photon gates will also affect the performance of the qubit registers. The photon loss can be directly absorbed into the overall photon transmission probability as previously described. Imperfect spin-photon gates and photonic switching affect the quality of the generated Bell pairs in a non-trivial way, which is treated in detail in Ref.~\cite{Xiaoyu2024}. In particular, the qudit encoding can result in correlated errors between the Bell pairs, which is a clear distinction between the qubit and the qudit mode of operation considered here. As shown in Ref.~\cite{Xiaoyu2024}, the effect of such correlated errors depends on the application; purification schemes are able to effectively target such correlated errors while quantum error correction codes are less effective. Since the focus of this work is to study the effect of limited coherence time of the qubit registers as well as the approximate entangled pair source, we will assume that these are the dominant errors and neglect the effect of other imperfections in the qubit register storage in our further analysis. This will allow us to provide general, hardware-agnostic bounds on the performance of both the qubit and qudit operation of a satellite-based SPDC source for long-distance entanglement distribution.         
 
\section{Performance} 

To compute the rate of entanglement distribution, we need to estimate the rate of photons successfully mapped to the qubit registers at the ground stations in both the qubit and qudit mode of operation. However, for the following analysis, the qubit mode of operation can be viewed as a special example of a 2-dimensional qudit. 

The length of a photonic time-bin is set by the repetition rate of the pump laser, $r_{\text{rep}}$. Consequently, an approximate qudit pair of the form in Eq.~(\ref{eq:Full_qudit_state}) is emitted at a rate of $r_{\text{rep}}/2^m$. If a detector clicks at a ground station, which ideally heralds a successful storage, the receiver will close the access to the qubits involved and send a (classical) heralding signal to the other ground station. If no signal has been received from the other ground station after the heralding time $\tau_h$ has elapsed (time of communication between the ground stations), the qubits will be immediately reset in order to receive photons again. Additionally, if a heralding signal for the corresponding qubits at the other ground station is received within the communication time, the qubits will also be immediately reset. A successful entanglement will only happen when the qubit pairs are available in both ground stations and heralding clicks are simultaneously recorded so that the ground stations both receive a heralding signal at the same time. Thus, the probability of successful entanglement distribution is given by 
$p_{\text{ent}} = p_{\text{suc}} \pi_{(0,0)}$,
where $ p_{\text{suc}}$ denotes the probability that a single qudit pair is successfully transmitted and mapped onto the qubit registers, assuming the qubits are available, and $\pi_{(0,0)}$ represents the probability that the qubits are indeed available. We calculate the latter probability using a Markov chain~\cite{Jones2013ADots}.

To increase the entanglement distribution rate, we consider a multiplexing scheme, where there are more available quantum memories than the desired number of entangled pairs. In particular, if $m$ entangled pairs are desired, we define a multiplexing parameter $n\in\{0,1,\ldots\}$ that specifies that $mn$ qubit memories are available at each ground station. We will count one entanglement generation attempt as attempting all $mn$ qubit memories sequentially. Thus, the time of one attempt is $\tau_c=2mn/r_{\text{rep}}$ ($\tau_c=2^{m}n/r_{\text{rep}}$) for the qubit (qudit) mode of operation. Additionally, we are in the limit $p_{\text{suc}}\ll1$ and we will therefore consider a slightly sub-optimal protocol where we only allow for a single heralding click per attempt. In other words, we discard events with more than a single heralding click per attempt. However, as we detail below, the probability of multiple heralding clicks is very low for practically relevant parameters meaning that including such events would not significantly increase the rate. 

We estimate the average rate of the entanglement distribution from the average number of attempts, $\langle A \rangle$ to obtain the desired number of event-ready entangled pairs:

\begin{equation}
    R = \frac{1}{\langle A \rangle \tau}.
\label{eq:rate}
\end{equation}

To calculate $\langle A \rangle$, it is useful to define the parameters $D$ and $N$ to keep track of the many possible combinations of obtaining $m$ entangled pairs across multiple attempts. In the qubit (qudit) mode of operation $D=m n$ ($D=n$) and $N=m$ ($N=1$). Since we have at most one heralding click per attempt, the average number of attempts to get $m$ entangled pairs can be estimated as:

\begin{equation}
\begin{aligned}[b]
\langle A \rangle &= \frac{D!}{(D-N)!\,p_{\text{approx}}}p_{ent}^N\sum_{i_1=0}^\infty \cdots\sum_{i_{N}>i_{N-1}}^\infty\bigl( i_1+1 + \\
& +\sum_{k=2}^N(i_k-i_{k-1})\bigr)(1-p_{ent})^{(D-N+1)i_N+D-N+\sum_{k=1}^{N-1}i_k}= \\
&= N+\sum_{k=1}^N \frac{(1-p_{ent})^{D+1-k}}{1-(1-p_{ent})^{D+1-k}}.
\label{eq:rate_multi}
\end{aligned}    
\end{equation}

This expression is valid both in the qubit and qudit mode of operation with the suitable choices of $D$ and $N$. It captures that in each attempt $i_j$, we go through all the available memories. As the probability of entanglement is small, we discard the events in which more than one pair is entangled in a single attempt. To account for all possible combinations where there are no simultaneous successes, we sum over these combinations with condition $i_k>i_{k-1}$ to ensure that each pair needs a different number of attempts. The term $1-p_{\text{approx}}$ is the probability of obtaining $m_t$ entangled pairs with more than one heralding click per attempt. This can be calculated as:

\begin{equation}
\begin{aligned}[b]
    p_{\text{approx}}& = \frac{D!}{(D-N)!}p_{\text{ent}}^N(1-p_{\text{ent}})^{ND-\left[(N-1)^2+2\right]}\\
    &\sum_{i_1=0}^{\infty} \cdots \sum_{i_N=0}^{\infty} (1-p_{ent})^{\sum_{k=1}^N\left(D + 1 - k\right)}.
\end{aligned}
\end{equation}

In our simulations, we keep $1-p_{\text{approx}}<1\%$, which ensures that we do not significantly reduce the rate by discarding events with multiple heralding clicks in one attempt. 

To compute the average fidelity of the entangled pairs, we take into account the effects of decoherence during the waiting period for all pairs to become successfully entangled.

\begin{equation}
    \begin{aligned}[b]
        \langle F(t)\rangle&= \frac{D!}{N\cdot(D-N)!p_{\text{approx}}}p_{ent}^N(1-p_{ent})^{ND-\left((N-1)^2 + 2\right)} \\
        &\sum_{i_1=0}^\infty\cdots\sum_{i_{N}=0}^\infty \Bigl[(1-p_{ent})^{\sum_{k=1}^{N}(D+1-k)i_k} \\
        &\bigl(F(\tau_h)+F(\tau_h+(i_{N}+1)\tau_c)+\cdots\\
        &+F(\tau_h+ \tau_c\sum_{j=2}^{N}(i_j+1))\bigr)\Bigr],
        \label{eq:fid_multi}   
    \end{aligned}
\end{equation}

where $F(t)$ is the fidelity of a pair after a storage time $t$. The detailed expression for $F(t)$ can be found in Appendix~\ref{app:fid}, which also includes infidelities from higher-order photon terms and dark counts. Note that we assume a worst-case scenario where these events result in states with fidelity of zero with the desired Bell states. The ordering used in Eq.~(\ref{eq:fid_multi}) means that  the $i_N$'th pair is the last pair that gets entangled and consequently, it only decoheres for the heralding time $\tau_h$. On the other hand, the first pair ($i_1$) has to wait the longest, namely for $\tau_h + \tau_c\sum_{k=2}^Ni_k$, where $i_k$ is the number of entanglement attempts needed for the $k^{th}$ pair to be successfully entangled.

The previous expressions for the average rate and fidelity of the entangled pairs (Eqs.~\eqref{eq:rate_multi},\eqref{eq:fid_multi}), do not operate with a cut-off time. In the qubit scheme, we add a cut-off time to the memories to ensure a certain target fidelity of the entangled pairs. We consider a strategy where once the first pair has been entangled, it will be stored in the quantum memory the chosen cut-off time, and the rest of the pairs we want to entangle must be successfully entangled within that time. To account for that in our expression of the rate and the fidelity, the sums over the attempts that occur after the first successfully entangled pair will be bounded by the cut-off time. The updated expressions including this can be found in the Appendix~\ref{app:A} for the case of $m=2$.

\begin{figure*}[htp]
   \centering
    \includegraphics[width=0.95\linewidth]{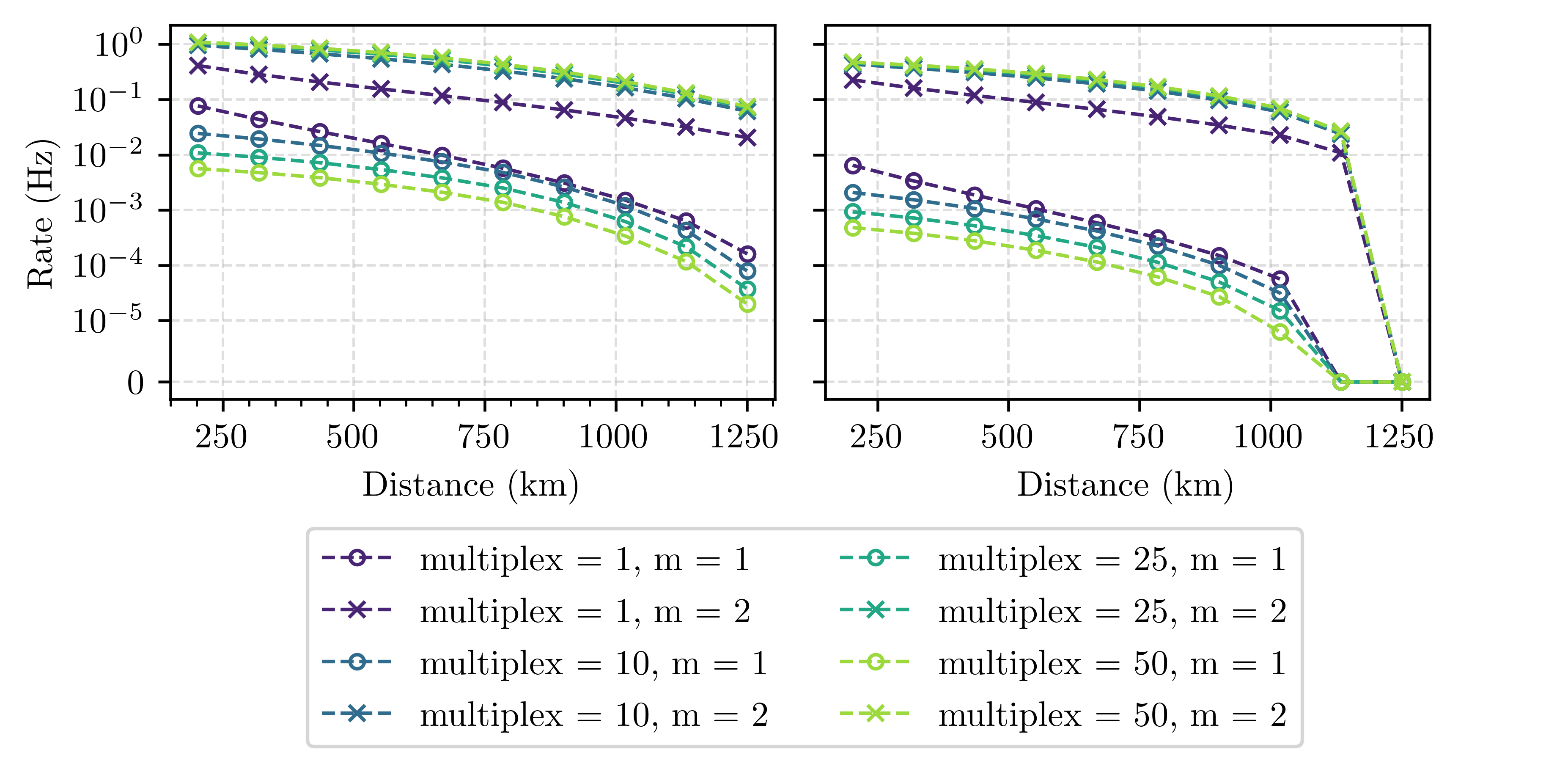}
    \caption{Performance comparison of the qudit (crosses) and qubit (circles) modes of operations. We optimize the squeezing parameter, $\lambda$, between values ranging from 0 to 0.1 as we are working in the weak-pump regime,  for each distance, as well as the cut-off time for the qubit case. We assume a dark count probability of $p_{\text{dark}}=1.6\cdot 10^{-5}$, a memory storage efficiency of $\eta=0.5$ (including photon detection), a SPDC repetition rate of $r_{\text{rep}}=10^{-7}$s, and a qubit decoherence time of $T_d = 10$s. Additionally, the transmission probability, $p_T$ is of the order of $10^{-3}$, being $\sim 8\cdot 10^{-3}$ for the shortest distance of $\sim 200$ km, and being $\sim 2 \cdot 10^{-3}$ for the longest distance of $\sim 1200$ km. In the \textbf{left figure}, the target fidelity is 0.9. The optimized value of the cut-off time is 1 s. In the \textbf{right figure}, the target fidelity is 0.95. Here, the optimized $t_{\text{cut}}$ is always 0.1 s.}
    \label{fig:fig_3}
\end{figure*}

In Fig.~\ref{fig:fig_3}, we show a comparison between the qudit and the qubit mode of operation when requiring a minimum average Bell pair fidelity of 0.90 (left figure) and 0.95 (right figure) of two event-ready entangled pairs (m=2). For the qubit mode of operation we optimize the cut-off time to achieve the maximum distribution rate for a certain distance. We assume the same performance in terms of memory efficiency and coherence time for both modes of operation and optimize the squeezing parameter $\lambda$ to achive the highest distribution rate. Our results show that the qudit mode of operation results in orders of magnitude larger distribution rates than the qubit mode. We also see that multiplexing increases the distribution rate for the qudit case while it surprisingly does not for the qubit mode of operation. This is because the attempt time increases as we increase the factor of multiplexing and the increase in success probability is not significant enough to outweigh this. In other words, multiplexing is worth implementing, if the rate at which the photons arrive at the ground stations, is higher than the communication time between them, which is not the case for the Micius parameters considered here where the SPDC source rate is 10 MHz. Note that increasing the repetition rate of the SPDC source would eventually result in a rate boost from multiplexing also in the qubit case.

Additionally, from Fig.~\ref{fig:fig_3}, we see that the use of $2^m$-dimensional qudits, where the dimensionality corresponds to the desired number of entangled pairs, enables entanglement over remarkably long distances of up to $1250$ km. However, we see that, at this maximum distance, the entanglement rate drops to zero when the target fidelity is $0.95$. The fidelity depends not only on the coherence time and the waiting time for the successful entanglement of all pairs, but also on the squeezing parameter ($\lambda$), the transmission probability ($p_T$), the memory efficiency ($\eta$), and the dark count probability ($p_{\text{dark}}$). In the simulations, the squeezing parameter is optimized to achieve the highest rate under the chosen fidelity cap. However, for distances $\geq 1250$ km, it becomes impossible to find a squeezing parameter that ensures a target fidelity of $0.95$ due to dark counts. This can be improved by either lowering the dark count rate of the detectors or improving the overall transmission probability. The transmission probability is influenced by factors such as the satellite's altitude and the size of its transmitters. Detection efficiency, on the other hand, can be enhanced by improving the detectors at ground stations. Therefore, while maintaining the same satellite, upgrading the ground station's detection systems could enable entanglement over even greater distances.

\section{Conclusion}

In summary, we have shown how satellite-assisted entanglement distribution can benefit significantly from adopting a qudit mode of operation when mutliple, high-fidelity, and event-ready entangled pairs are desired.
Our simulations show that a qudit mode of operation results in several orders of magnitude higher distribution rates than a standard qubit mode of operation for all distances between 200-1200km. Importantly, the qudit mode of operation does not require any substantial upgrades of existing quantum satellites payloads but rather upgrades of the quantum memories at the ground nodes, which are much more accessible. To achieve a general bound on the achievable rate, we have assumed negligible infidelities from the memory storage operation, except for the effect of limited coherence time. Although errors in current hardware do not quite satisfy this assumption for the target fidelities of 90\% and 95\% assumed in this work, we note that recent experiments with cavity-coupled diamond color centers have demonstrated spin-photon gates with infidelities of $\sim 2\%$, qubit gate errors of $\sim 1\%$ and measurement errors of $\sim 1-2\%$~\cite{wei2024} together with single photon storage efficiencies of~$\sim 40\%$~\cite{Bhaskar2020ExperimentalCommunication} with clear paths to improvement. Additionally, using nearby nuclear memory spins, coherence times on the order of seconds have been demonstrated with the same hardware~\cite{Nguyen2019,Stas2022}, with the possibility to extend the coherence time to the order of minutes using weakly coupled nuclear spins~\cite{Bradley2019}. We also note that Ref.~\cite{wei2024} reports entanglement generation between two SiV qubit memories using a photonic qudit demonstrating the compatibility of this hardware with photonic qudits. This makes it conceivable that satellit-assisted high-quality entanglement distribution using the qudit mode of operation considered in this work could be realized with near-term hardware.  
\begin{acknowledgements}
We acknowledge helpful discussions with Kenneth Goodenough and Francisco Ferreira da Silva . We acknowledge funding from the NWO Gravitation Program Quantum Software Consortium (Project QSC No. 024.003.037). J.B. acknowledges support from The AWS Quantum Discovery Fund at the Harvard Quantum Initiative. \newline \newline
\textbf{Data Availability:} The code and data of the results of this paper are openly available on 4TU.ResearchData: "Data underlying the publication "Satellite-assisted Entanglement Distribution with High-Dimensional Photonic Encoding", at \url{https://doi.org/10.4121/61ca8a5b-3af2-4947-904c-36c21b4ed090}, Ref.\cite{dataset}.
\end{acknowledgements}

\appendix
\section{Fidelity}\label{app:fid}
The fidelity of each entangled pair at the ground stations is defined as:

\begin{equation} \label{eq:fidel1}
    F(t) = \frac{\langle \Phi^+| \rho_{\text{pair}}|\Phi^+\rangle}{Tr(\rho)}, 
\end{equation}

where:

\begin{eqnarray}
    \langle \Psi^+|\rho_{\text{pair}} | \Psi^+\rangle &=& 2^mp_T^2\eta^2\lambda^2((e^{-t/T_A}+(1-e^{-t/T_A})\epsilon_1)\nonumber \\
    &&(e^{-t/T_B}+(1-e^{-t/T_B})\epsilon_1)
    +(1-e^{-t/T_A})\nonumber \\
    &&(1-e^{-t/T_B})(\epsilon_x^2+\epsilon_y^2+\epsilon_z^2)),
\end{eqnarray}

\begin{eqnarray}
    Tr(\rho) & = & 2^m p_T^2\eta^2\lambda^2+2^{m-1}(2^m+1)p_T^4\lambda^4\eta^4 \nonumber \\
    && + 2^{m+1}p_T\eta(1-p_T\eta)\lambda^2p_{\text{dark}} +2^m(2^m+1)p_T^2\eta^2 \nonumber \\
    && (1-p_T\eta)^2\lambda^4\Bigl(p_{\text{dark}}+2\Bigr) +2^{m+1}(2^m+1)\nonumber\\
    && \Bigl(p_T^3\eta^3(1-p_T\eta)\lambda^4+p_T\eta(1-p_T\eta)^3\lambda^4p_{\text{dark}}\Bigr) \nonumber \\
    &&+p_{\text{dark}}^2\Bigl(2^{m-1}(2^m+1)(1-p_T\eta)^4\lambda^4\nonumber \\
    && + 2^m(1-p_T\eta)^2\lambda^2+1\Bigr)+ \mathcal{O}(\lambda^6),
\end{eqnarray}

\begin{equation}
    |\Phi^+\rangle = \frac{1}{\sqrt{2}}\left(|0\rangle|0\rangle + |1\rangle|1\rangle\right).
\end{equation}

Here, $|\Phi^+\rangle$ is the desired Bell pair, and $\rho_{\text{pair}}$ is the (unnormalized) density matrix of a single pair of entangled memory qubits. In the qudit mode of operation, the latter corresponds to tracing out all other qubit pairs. The normalization of $\rho_{\text{pair}}$ corresponds to the probability of obtaining successful heralding clicks that prepare the memories in this state. The denominator in Eq.~(\ref{eq:fidel1}) is the total probability of a successful heralding, which can be calculated as the trace over the full (unnormalized) density matrix, $\rho$ including contributions from higher order photon terms and dark counts. Note that we adopt a lower bound on the fidelity by assuming that dark counts and higher-order photon terms results in states with zero overlap with the desired Bell states. 

As stated in the main text, $p_T$ is the total transmission probability and we explicitly include $\eta$ as the memory storage efficiency. The time the pair is stored before successful entanglement of all desired pairs is achieved is $t$. Note that this varies from pair to pair in the qubit mode of operation as shown in Eq.~(\ref{eq:fid_multi}), while $t=t_h$ in the qudit mode of operation. We let $T_A$ and $T_B$ denote the coherence times of Alice and Bobs memory qubits and let $\epsilon_1$,$\epsilon_x$, $\epsilon_y$, and $\epsilon_z$ represent the relative error rates for Pauli errors in the I, X, Y, and Z bases, respectively. In our simulations, we set both decoherence times to $T_A = T_B = \tau_{\text{coh}}$ and assume a depolarizing channel with $\epsilon_1=\epsilon_x=\epsilon_y=\epsilon_z=1/4$.

\section{Cut-off time calculation.} \label{app:A}

Here, we show how we estimate the rate and fidelity for the qubit mode of operation when we have cut-off time and we want to entangle two pairs. In this scheme, the second pair needs to be entangled within the cut-off time, otherwise, the first entangled pair is discarded and we repeat the process again. To estimate the average number of attempts to successfully entangle two pairs within the cut-off time, we consider the average number of tries assuming a successful entanglement generation i.e. that the pairs are generated within the cut-off time, $\langle n_{\text{succ}}\rangle$. The expression for this is similar to the one shown in Eq.~\eqref{eq:rate_multi}, except that the sum over the second pair, $i_2$, is bounded by a finite number of tries given by the cut-off time, $N_{\text{cut}}$. This number of tries is defined as $N_{\text{cut}}=t_{\text{cut}}/D\tau_c$, where $t_{\text{cut}}$ is the cut-off time, which is always smaller than the decoherence time of the quantum memories. Specifically, 

\begin{equation}
\begin{aligned}[b]
    \langle n_{\text{succ}} \rangle & = \frac{D!}{(D-2)!p_{\text{2p}}}p_{\text{ent}}^2(1-p_{\text{ent}})^{(2D-3)}\\
    &\sum_{i_1=0}^{\infty}\sum_{i_2=0}^{N_{\text{cut}}-2}\bigl(i_1 + i_2 + 2\bigr)(1-p_{\text{ent}})^{i_2(D-1) + Di_1},
\end{aligned}
\end{equation}

where $p_{\text{2p}}$ is given by

\begin{equation}
\begin{aligned}[b]
    p_{\text{2p}} &= \frac{D!}{(D-2)!}p_{\text{ent}}^2(1-p_{\text{ent}})^{(2D-3)} \nonumber \\
    &\sum_{i_1=0}^{\infty}\sum_{i_2=0}^{N_{\text{cut}}-2}(1-p_{\text{ent}})^{i_2(D-1) + Di_1},
\end{aligned}
\end{equation}
and is the probability of successfully entangling two pairs within the cutoff time. Note that we still only allow for a single success per attempt as described in the main text. To estimate the average number of attempts, we also estimate the the average number of tries,  $\langle n_{\text{fail}}\rangle$., to obtain one entangled pair but fail to entangle the second one within the cut-off time as well as the probability for this event. This is given by

\begin{equation}
\begin{aligned}[b]
    \langle n_{\text{fail}}\rangle & =  N_{\text{cut}}+\langle n_1 \rangle = \nonumber \\
    &\frac{D!}{(D-1)!}p_{\text{ent}}(1-p_{\text{ent}})^{D-2}\sum_{i_1 = 0}^{\infty}\bigl(i_1 + 1\bigr)(1-p_{\text{ent}})^{Di_1} \nonumber \\
     &+ N_{\text{cut}},
\end{aligned}
\end{equation}
where $\langle n_1 \rangle$ is the average number of attempts to get the first pair. The probability of a failed attempt is $p_{\text{fail}}=(1-p_{\text{ent}})^{N_{\text{cut}}}$.

We now estimate the average number of attempt to successfully entangle two pairs within the cutoff time as 
\begin{equation}
\begin{aligned}[b]
    \langle n \rangle& = \sum_{i=0}^{\infty}\bigl( i \langle n_{\text{fail}}\rangle + \langle n_{\text{succ}}\rangle\bigr) p_{\text{fail}}^ip_{\text{succ}} =\\
    &\frac{1}{p_{\text{succ}}}\bigl(\bigl(\langle n_{\text{fail}} \rangle - \langle n_{\text{succ}} \rangle \bigr)p_{\text{fail}} + \langle n_{\text{succ}}\rangle\bigr),
\end{aligned}
\end{equation}

where $p_{\text{succ}}= 1- p_{\text{fail}}$. We note that $p_{\text{succ}}\approx p_{\text{2p}}$ for the parameters considered since the probability to heralding two pairs in the same attempt is very small. For the computation of the fidelity, we also truncate the upper limit of the sum for $i_2$.

\begin{equation}
\begin{aligned}[b]
\langle F(t) \rangle &= \frac{D!}{2(D-2)!p_{\text{2p}}}p_{\text{ent}}^2(1-p_{\text{ent}})^{2D-3} \nonumber\\
&\sum_{i_1=0}^{\infty}\sum_{i_2=0}^{N_{\text{cut}}-2} \bigl(1-p_{\text{ent}}\bigr)^{i_2(D-1) + Di_1}\bigl[F(\tau_h)\nonumber \\
&+F(\tau_h + (i_2 +1)\tau_c\bigr]
\end{aligned}
\end{equation}

\bibliography{main.bib}

\end{document}